\documentclass[aps, pra, onecolumn, notitlepage, 12pt, amsmath, amssymb,showkeys]{revtex4-1}

\usepackage{braket}
\usepackage[pdftex]{graphicx} 
\usepackage{color}

\begin{document}
\title{Revisiting Quantum Feedback Control: Disentangling the Feedback-induced Phase from the Corresponding Amplitude}

\author{Kisa Barkemeyer}
\author{Regina Finsterh\"olzl}
\author{Andreas Knorr}
\author{Alexander Carmele}
\thanks{alex@itp.tu-berlin.de}
\affiliation{Institut f\"ur Theoretische Physik, Technische Universit\"at Berlin, D-10263 Berlin, Germany}

\begin{abstract}
Coherent time-delayed feedback allows the control of a quantum system and its partial stabilization against noise and decoherence. The crucial and externally accessible parameters in such control setups are the round-trip-induced delay time $\tau$ and the frequencies $\omega$ of the involved optical transitions which are typically controllable via global parameters like temperature, bias or strain. They influence the dynamics via the amplitude and the phase $\phi = \omega \tau$ of the feedback signal. These quantities are, however, not independent. 
Here, we propose to control the feedback phase via a microwave pump field. Using the example of a $\Lambda$-type three-level system, we show that the Rabi frequency of the pump field induces phase shifts on demand and therefore increases the applicability of coherent quantum feedback control protocols.
\end{abstract}

\maketitle

\section{Introduction}
The processing of quantum information relies on three principal components: the storage, the manipulation, and the transmission of quantum information. 
While quantum few-level systems such as atoms or molecules can be used for information storage and manipulation, photons are the ideal candidates for carrying quantum information from one point to another. Thus, growing interest in the field of quantum computation and quantum information on the part of science as well as industry calls for the possibility to efficiently control quantum optical systems \cite{Nielsen2010, Scully1997, Cirac1997, Petersen2010}.
An important task is to prevent the loss of information due to decoherence and noise which is crucial for the successful implementation of quantum information processing \cite{Zoller2005}. 
Several protocols have been proposed which include quantum error correction, quantum gate purification, and entanglement purification schemes \cite{Shor1995, VanEnk1997, Bennett1996, Vermersch2017}.

Another promising approach is coherent quantum feedback control.
Feedback has long been a common means for the control of classical and semiclassical systems \cite{Pyragas1992, Bechhoefer2005, Scholl2008, Scholl2016, Loos2019}. Examples include semiconductor laser setups and chemical systems like the Belousov-Zhabotinsky reaction \cite{ Lang1980, Schneider1993, Janson2004, Otto2012, Schulze2014, Munnelly2017}. 
In the quantum regime, coherent feedback control was first introduced as all-optical feedback by Wiseman and Milburn \cite{Wiseman1994} and is based on a fully quantum mechanical control mechanism \cite{Lloyd2000}. 
It has been successfully implemented in a number of experimental setups \cite{Nelson2000, Iida2012, Kerckhoff2012, Kerckhoff2013, Hirose2016}.
Such schemes allow the preservation of coherence as opposed to measurement-based feedback control where repeated measurements result inevitably in the destruction of coherence \cite{Steixner2005, Wiseman2009, Serafini2012, Yamamoto2014, Jacobs2014, Zhang2017, Sudhir2017, Sayrin2011, Zhou2012}.
If the feedback delay time $\tau$, that is the time between the emission of a signal and the reabsorption from a reservoir, is not negligible, partial quantum entanglement is preserved. As a consequence, the non-Markovianity of the dynamics needs to be taken into account \cite{Giovannetti1999, Tufarelli2014, Grimsmo2015, Albert2011, Hoi2015,deVega2017,Carmele2019,whalen2016time,calajo2019exciting,PhysRevA.91.053845}. %

Various setups to control quantum few-level systems via time-delayed feedback have been studied theoretically and it has been shown that it is possible to control characteristic quantities such as the photon-photon correlation and the concurrence which functions as a measure of entanglement
\cite{Dorner2002, Schaller2012, Carmele2013, Hein2014, Kabuss2016, Pichler2016, Nemet2016, Lu2017, Metelmann2017, Guimond2017, Nemet2019, Droenner2019, Calajo2019}. 
In these systems, in general, the control parameters that can be used to evoke the desired behavior are the delay time $\tau$ and the characteristic frequency $\omega$ which, depending on the considered setup, can be e.g. the frequency of an involved optical transition or the frequency of a cavity mode.
These parameters influence the dynamics via the feedback amplitude $x(t-\tau)$ and the phase $\phi = \omega \tau$.  Tuning them, however, requires major rearrangements in the setup. Furthermore, the quantities are intertwined since changes of the delay time $\tau$ influence both the amplitude and the phase of the feedback signal.

In this Paper, we propose a novel approach for the non-invasive and instantaneous control of few-level systems through the application of an external pump field to tune the phase between incoming and emitted signal.
Using the example of a $\Lambda$-type three-level system (3LS) subjected to coherent time-delayed feedback we study how an additional pump field, resonant to the transition between the non-degenerate ground states, influences the dynamics and gives rise to a new control parameter, the Rabi frequency of the field. With this control scheme, it is possible to disentangle the control of the feedback phase from the control of the system's transition frequencies and the delay time.

The paper is organized as follows. In Sec.~\ref{3LS} we motivate the study of the $\Lambda$-type 3LS for which the application of an external laser field yields the Rabi frequency as a new control parameter.
In Sec.~\ref{Rev} we then briefly present three existing setups from the literature employing coherent feedback control schemes for which the independent control of the phase is crucial.
This short review is given to stress the importance of the possibility to address the phase individually. It is demonstrated that the interplay of amplitude and phase has the potential to increase the degree of entanglement of emitted photon pairs in a biexciton cascade \cite{Hein2014}, to suppress or enhance bunching as well as antibunching in photon-correlation detection schemes \cite{Lu2017}, and to manipulate one- and two-photon processes in time-resolved resonance fluorescence experiments \cite{Droenner2019}.
If the previously studied systems are extended in a way that allows its application, our scheme potentially enables and simplifies the control of the emerging phenomena which are instanced to inspire new applications in the field of coherent quantum feedback \cite{Grimsmo2015,glaetzle_feedback,waks_feedback,guimond_feedback,Whalen2016,Nemet2016}. 
Subsequently, in Sec.~\ref{results} we derive our new control scheme in detail and compare it to the scheme in which only the delay time $\tau$ is used as a control parameter. We demonstrate that the application of a microwave pump field opens up new possibilities in potential experimental realizations and allows a fast and efficient stabilization of the excitation as well as population trapping \cite{Dorner2002,Nemet2019_arxiv,nemet_fiber}.

\begin{figure}
\centering
\includegraphics[width=0.35\linewidth]{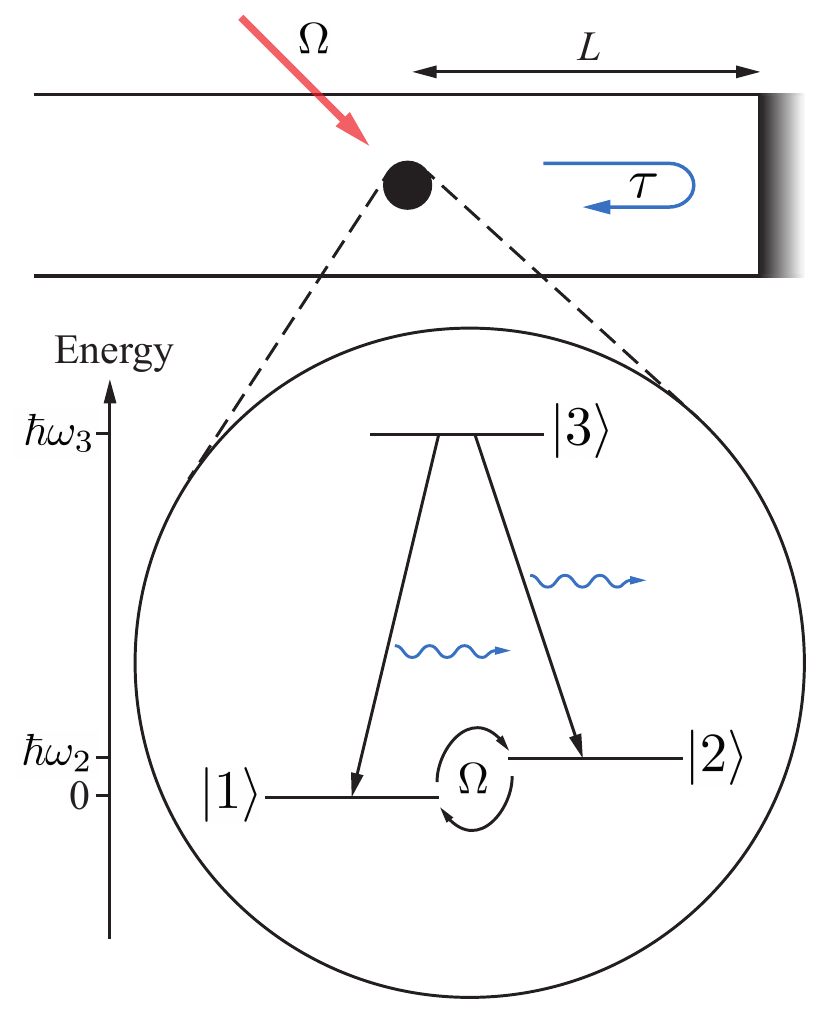}
\caption{The considered non-degenerate $\Lambda$-type three-level system (3LS) interacting with a photonic reservoir inside a semi-infinite one-dimensional waveguide. The edge of the waveguide at a distance $L$ from the 3LS acts as a mirror and thus provides feedback with delay time $\tau$. A pump field with Rabi frequency $\Omega$ is applied which is resonant to the $\ket{1} \leftrightarrow \ket{2}$ transition.}
\label{sys}
\end{figure}

\section{Proposal to non-invasively control the feedback phase in a $\Lambda$-type system}
\label{3LS}

We consider a $\Lambda$-type 3LS interacting with a photonic reservoir inside a semi-infinite one-dimensional waveguide. Radiative transitions are possible between the excited state $\ket{3}$ and one of the two non-degenerate ground states $\ket{1}$ and $\ket{2}$. Between the ground states, a resonant pump field is applied. The setup is depicted in Figure~\ref{sys} and the Hamiltonian of the system reads
\begin{align}
  \label{H}
 H(t) = &\phantom{+} \hbar \omega_2 \sigma_{22} + \hbar \omega_3 \sigma_{33} + \hbar \int \mathrm{d}k  \omega_k r_k^\dagger r_k 
 \notag \\ &+ \hbar \Omega(t) \cos\left( \omega_2 t\right) \left( \sigma_{12} + \sigma_{21}\right) + \hbar \int \mathrm{d}k \left[ g_k r_k^\dagger \left( \sigma_{13} + \sigma_{23} \right) + H.c. \right]. 
\end{align}
In this expression, the first line describes the non-interacting system, the first term in the second line models the external pumping and the last term results from the interaction between the 3LS and the reservoir. %
All energy values are considered relative to level $\ket{1}$. The energy of level $\ket{2}$ ($\ket{3}$) is $\hbar \omega_{2}$ ($\hbar \omega_3$) with the corresponding occupation number operator $\sigma_{22}$ ($\sigma_{33}$). The annihilation (creation) of a photon in reservoir mode $k$ with energy $\hbar \omega_k$ is described by the bosonic annihilation (creation) operator $r_k^{(\dagger)}$.
The atomic flip-operators $\sigma_{ij}$, $i, j = 1,2,3$, are defined as $\sigma_{ij} = \ket{i}\bra{j}$ and satisfy the commutation relation $
\left[ \sigma_{ij} , \sigma_{kl} \right] = \sigma_{il} \delta_{jk} - \sigma_{kj} \delta_{il}$.
The pump field is characterized by its pump frequency $\omega_2$ and its amplitude $\hbar \Omega$ with Rabi frequency $\Omega$. The pump frequency coincides with the transition frequency of level $\ket{1}$ and level $\ket{2}$ since we assume resonance. 
Because of the mirror at distance $L$ from the 3LS, an emitted signal is fed back into the 3LS after the delay time $\tau = 2L/c$ where $c$ is the speed of light.
We include the feedback mechanism into our calculations by assuming a non-Markovian environment, that is a structured reservoir which results in a sinusoidal dependence of the coupling strength $g_k$ on the photon mode $k$, so that $g_k = g_0 \sin(kL)$ \cite{Cook1987, Dorner2002, Carmele2013}. 

If we set $\Omega(t)\equiv 0$ in the Hamiltonian given in Eq.~\eqref{H}, the dynamics of the probability amplitude $c_3(t)$ which describes the excited state of the 3LS can be derived analytically \cite{Dorner2002,Kabuss2015,sinha2019non}. In the special case of $\omega_2 \tau = 2 \pi n$, $n \in \mathbb{N}$, this yields a Lambert $W$-function in Laplace space and in time domain we obtain
\begin{equation}
\dot{c}_3(t) = -2\Gamma \left[ c_3(t) -  e^{i\omega_3\tau}\theta(t-\tau)c_3(t-\tau) \right]
\end{equation}
with decay rate $\Gamma = g_0^2 \pi/(2 \hbar^2 c)$. Note that this is the same delay differential equation as the one we obtain for a two-level system under the influence of coherent feedback. 
Thus, the phase $\omega_3\tau$ determines whether the delayed amplitude leads to an accelerated or decelerated decay. The fastest decay occurs for $\omega_3\tau=(2n+1)\pi$, $n \in \mathbb{N}$, while the decay is maximally slowed down for $\omega_3\tau=2n\pi$, $n \in \mathbb{N}$.
However, if $\tau$ is the only free parameter and we assume fixed transition frequencies, the phase is automatically set with $\tau$ and not independent of the delayed amplitude.
If we, on the other hand, choose a time-independent microwave driving field, $\Omega \neq0$, we obtain for the dynamics of the probability amplitude
\begin{widetext}
\begin{multline}
\dot{c}_3(t) = - 2 \Gamma c_3(t) + \Gamma e^{i \omega_3 \tau} \left[\cos\left( \frac{\Omega}{2}\tau\right)\left(1+e^{-i\omega_2 \tau} \right) \right. \left. - i \sin \left( \frac{\Omega}{2}\tau \right) \left( e^{-i \omega_2 t }+ e^{i \omega_2 (t-\tau)}\right) \right] \\
\times c_3(t- \tau)  \Theta(t-\tau). \label{DDEPump}
\end{multline}
\end{widetext}
From the above equation we see that the time-independent pumping strength $\Omega$ now enters as an additional control parameter. Depending on the transition frequencies $\omega_2$ and $\omega_3$, the delay time $\tau$, and $\Omega$ different scenarios occur.
This equation is the main result of the paper.
By tuning $\Omega$, destructive and constructive feedback scenarios can be realized without the need to change the distance between the mirror and the 3LS.
This strongly simplifies the study of systems subjected to coherent quantum feedback and allows to unravel amplitude-based impacts from effects which rely on a well-controlled feedback phase.
It is also possible to assume time-dependent pumping, however, Eq.~\eqref{DDEPump} is only valid in the continuous-wave regime.
Before we derive this equation in detail, we motivate why the control of the phase independent of the amplitude is of great importance for the realization of proposed quantum-feedback phenomena.

\section{Conventional coherent feedback control schemes}
\label{Rev}

Various systems under the influence of coherent quantum feedback have been studied and it has been shown that in this way the control of a wide range of phenomena is possible. To evoke a desired behaviour, in many cases, a specific delay time as well as a certain feedback phase is essential. However, in most of the proposed setups, the delay time is the only accessible parameter which influences the dynamics via the phase as well as the amplitude of the feedback signal. In this section we briefly discuss common implementations of coherent feedback control schemes to set the scene for the presentation of our approach in which the Rabi frequency of an external pump field arises as a new control parameter and allows the individual control of the feedback phase.

\subsection{Selective control of individual photon probabilities}
\label{2LS}

\begin{figure}
\centering
\includegraphics[width=7.5cm]{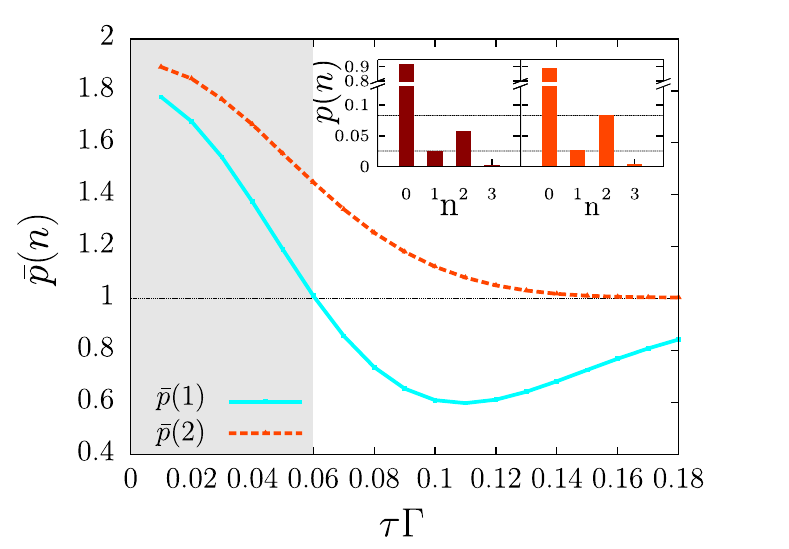}
\caption{Photon emission probabilities $\bar{p}(n)$, $n= 1,2$, normalized with respect to the case without feedback ($\bar{p}(n)=1$) as functions of the delay time $\tau$ scaled with the decay rate $\Gamma$ for destructive feedback, i.e. $\phi=(2n+1)\pi$, $n \in \mathbb{N}$. For pulse area $A=2\pi$ we find that the two-photon probability $\bar{p}(2)$ (red, dashed line) is enhanced by feedback while the single-photon probability $\bar{p}(1)$ (blue, solid line) can be either enhanced or suppressed. Inset: Photon emission probability $p(n)$, $n=0,1,2,3$, without feedback (red, left) and with feedback at $\tau \Gamma = 0.06$ (orange, right) where the two-photon probability is increased by about 50\%. \copyright ~2019 American Physical Society, reprinted from \cite{Droenner2019}.}
\label{Results_Droenner}
\end{figure}

The first setup we consider is a two-level system (2LS) in front of a mirror examined in, including, but not limited to, \cite{Droenner2019,Grimsmo2015,Pichler2016,Dorner2002, Kim2014, Fang2015, Whalen2016}.
The 2LS is pumped with a pulsed laser field $\Omega(t)$ which controls the emission statistics via its pulse area $A$. 
Looking at the emission without feedback, we observe that for a pulse with $A = \pi$ the 2LS is inverted and acts as a single-photon source. On the contrary, for a pulse with $A = 2\pi$ the possibility that two photons are emitted dominates over the one of a single-photon event \cite{Fischer2017}. It can be shown that it is possible to enhance or suppress individual photon probabilities using time-delayed feedback since they respond differently to the applied control of the Pyragas type \cite{Droenner2019}.
For an exemplary operator $x(t)$ in the Heisenberg picture we obtain a differential equation of the form
\begin{equation}
 \dot{x}(t) = \frac{i}{\hbar} \left[H_S, x(t) \right] - K\left[ x(t) - e^{i\phi} x(t-\tau)\right] + N(t). 
\end{equation}
Here, $H_S$ denotes the Hamiltonian of the system. $K$ is a control force which vanishes when a steady state or periodic orbit is reached. The noise contribution $N(t)$ ensures that the canonical commutation relations of the system operators are conserved.
The control parameter that is tuned to achieve the desired behavior in this case is the delay time $\tau$. 
With the delay time, a certain feedback phase $\phi = \omega_0 \tau$ is set where $\omega_0$ is the transition frequency of the 2LS. It determines whether the interference between emitted and fed back signal is constructive or destructive.
The feedback phase $\phi$ is a crucial parameter which arises when the classical Pyragas control scheme is extended to the quantum regime. 
Depending on the feedback phase $\phi$, the excitation decay from the 2LS is either accelerated or slowed down compared to the free decay in the Wigner-Weisskopf model. When a pump pulse of area $A=2\pi$ is applied, the two-photon possibility can be enhanced whereas the possibility that a single photon is emitted can be enhanced or suppressed by applying destructive feedback, that is feedback with a phase $\phi = (2n+1)\pi$, $n \in \mathbb{N}$. This can be seen in Figure~\ref{Results_Droenner}. One realizable scenario is an enhancement of the two-photon probability of around 50\% and a simultaneous preservation of the single-photon probability if the delay time $\tau$ and phase $\phi$ is chosen appropriately. 

\subsection{Intensified antibunching}

\begin{figure}
\centering
\includegraphics[width=7.5cm]{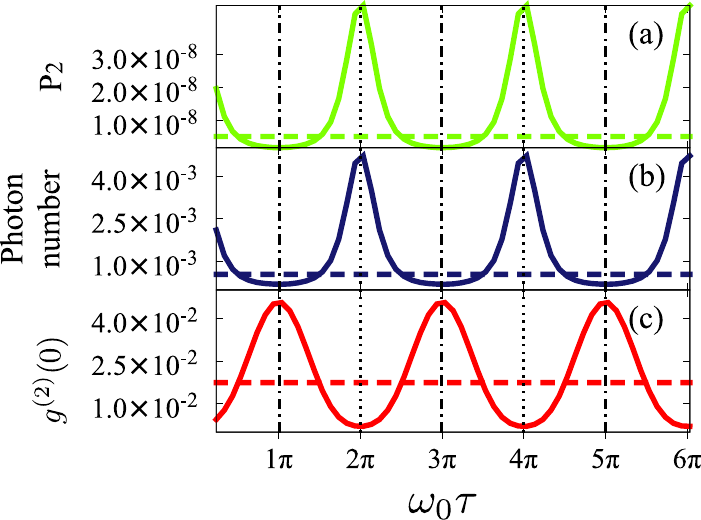}
\caption{Effect of feedback on the system in which antibunching can be observed (solid lines) in comparison to the case without feedback (dashed lines). Shown are (a) the two-photon probability $p_2$, (b) the photon number, and (c) the second order correlation function $g^{(2)}(0)$ as functions of the feedback phase $\phi = \omega_0 \tau$. While a (destructive) feedback phase of $\phi = (2n+1) \pi$, $n \in \mathbb{N}$, (dot-dashed lines) leads to low intensity and the suppression of antibunching, a (constructive) feedback phase $\phi = 2\pi n$, $n \in \mathbb{N}$, (dotted lines) causes enhanced antibunching and high intensity. \copyright ~2017 American Physical Society, adapted from \cite{Lu2017}.}
\label{Results_Lu}
\end{figure}

Another setup for coherent feedback control consists of two 2LS interacting via the respective coupling $g_i$ with a single-mode cavity \cite{Lu2017}. The cavity is pumped by a weak continuous-wave pump field $\Omega \ll g_i$ and placed inside a semi-infinite photonic waveguide which provides non-Markovian feedback at delay time $\tau$.
The evaluation of the second-order correlation function for zero time delay, $g^{(2)}(0)$, allows drawing conclusions about the photon statistics of the light emitted from the cavity. A value smaller than one, i.e. $g^{(2)}(0)<1$, corresponds to a non-classical state of the light field where the probability to measure two photons simultaneously is decreased compared to a coherent laser field, i.e. the quantum emission is antibunched. 
Without feedback, the system shows antibunching if a suitable detuning $\Delta$ between the frequency of the cavity mode and the external pump frequency is chosen.
It is shown that this effect can be enhanced or suppressed significantly by applying feedback, cf. Figure~\ref{Results_Lu}.  The influence of the feedback is  
determined by the feedback phase $\phi= \omega_0 \tau$ with cavity mode frequency $\omega_0$ and delay time $\tau$. This phase can be tuned to control the emission statistics.
In Figure~\ref{Results_Lu}, the values of the two-photon probability $p_2$, the photon number, and the second-order correlation function $g^{(2)}(0)$ oscillate with $\phi$. Extreme cases occur for destructive feedback, where $\phi = (2n+1)\pi$, $n \in \mathbb{N}$, and constructive feedback, where $\phi = 2\pi n$, $n \in \mathbb{N}$. In the destructive case, the two-photon probability and the photon number show a minimum while the $g^{(2)}(0)$ function exhibits a maximum. Its value lies above the one that is obtained without feedback, i.e. antibunching is reduced. On the contrary, if constructive feedback is applied, the maximal two-photon probability and photon number can be observed while the $g^{(2)}(0)$ function shows a minimum that lies below the no-feedback value. From this, we can conclude that an enhancement of the antibunching is obtained for constructive feedback and at the same time the intensity of the emitted light is maximized. Since typically antibunching is reduced for higher photon numbers, e.g. $g^{(2)}(0) = 1 - 1/\left<n\right>$ for a Fock state, this counterintuitive feature is a clear signature of underlying quantum interference effects between system and reservoir states.

\subsection{Enhanced photon entanglement}\label{4LS}

The third setup we consider is a quantum dot (QD) biexciton cascade which is covered in \cite{Hein2014, Bounouar2015, Heindel2017, Bounouar2018}. In a radiative relaxation process, the initially excited QD emits either two horizontally ($H$) or two vertically ($V$) polarized photons because angular momentum is conserved. This leads to the formation of a Bell state of the form
\begin{equation}
    \ket{\psi} = \frac{1}{\sqrt{2}}\left(\ket{HH} + \ket{VV} \right).
\end{equation}
Thus, the QD biexciton cascade serves as an important source for polarization-entangled photons. The entanglement is, however, reduced if a finite excitonic fine-structure splitting $\hbar \delta$ is present, as it renders the relaxation paths distinguishable.

It can be shown that coherent feedback control can be used as a means to counteract this reduction of entanglement.
Depending on the feedback phase $\phi = \omega_X \tau$, where $\omega_X$ is the excitonic transition frequency and $\tau$ is the delay time, the impact of the feedback differs.

\begin{figure}
\centering
\includegraphics[width=9.5cm]{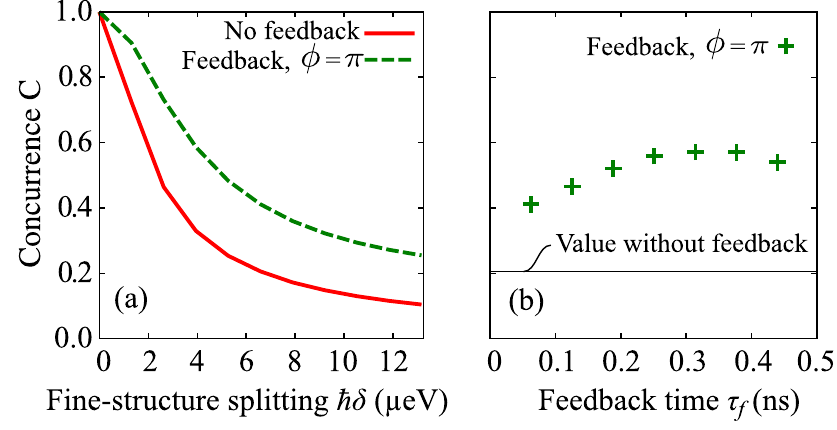}
\caption{Photon concurrence $C$ depending on different parameters. (a) Concurrence as a function of the finite fine-structure splitting $\hbar \delta$ without feedback (red, solid line) and with destructive feedback at phase $\phi = (2n+1)\pi$, $n \in \mathbb{N}$ (green, dashed line). The concurrence decreases with increasing fine-structure splitting but can be enhanced significantly by applying feedback. (b) Concurrence as a function of the feedback time $\tau$ at $\phi = (2n+1)\pi$, $n \in \mathbb{N}$ and $\delta = 10\text{\,ns}^{-1}$. For all delay times $\tau$ the concurrence is larger than without feedback, there is, however, an ideal delay time for which the concurrence is maximized. \copyright ~2014 American Physical Society, adapted from \cite{Hein2014}.}
\label{Results_Hein}
\end{figure}

The entanglement is measured via the concurrence $C$ of the reduced photonic density matrix.
As can be seen in Figure~\ref{Results_Hein}~(a), the concurrence and thus the entanglement decreases with increasing fine-structure splitting $\hbar \delta$. Nevertheless, when applying destructive feedback with a phase $\phi = (2n+1)\pi$, $n \in \mathbb{N}$, the entanglement can be enhanced considerably. For delay times $\tau$ small compared to the biexciton lifetime this can be understood by looking at the emission spectrum: Since destructive feedback leads to a faster decay of the excitation in the QD compared to the case without feedback, the emission peaks centered at $\omega_X \pm \delta/2$ in the spectrum are broadened. Due to the larger spectral overlap between the emission peaks, the decay paths are less distinguishable. Therefore, the entanglement is partially recovered despite the fine-structure splitting.

Figure~\ref{Results_Hein}~(b) shows that by varying the feedback time $\tau$ the concurrence and with it the amount of entanglement can be tuned. The entanglement is always larger than in the case without feedback, at a certain value, however, the entanglement is maximized. This confirms our initial statement that the delay time influences the dynamics not only via the feedback phase but also in other ways, e.g. via the amplitude of the feedback signal.

\section{Comparison of the approaches for coherent feedback control}
\label{results}

The examples presented above, Sec.~\ref{Rev}\,A--C, demonstrate that feedback can be used to coherently control a variety of systems and phenomena employing the system's past. However, the time-delayed amplitude of the signal $x(t-\tau)$ and the feedback phase $\phi=\omega \tau$, which determines how emitted and fed back signal interfere, are not independent in the above examples, assuming the system frequencies to be fixed.
Since the time delay $\tau$  appears in the amplitude as well as in the phase of the feedback signal, it is reasonable to look for possibilities to disentangle the control of the amplitude from the control of the phase. This can be achieved by tuning the respective transition or cavity mode frequency $\omega$, e.g. via bias, strain or temperature.
However, such control schemes typically introduce additional decoherence channels via temperature effects or additional fields.
Our novel approach of non-invasive and instantaneous feedback control via an additional external laser field can be an alternative tuning knob.

We now derive in detail the dynamics of the probability amplitude of the excited state as stated in Eq.~\eqref{DDEPump}.
To that end, we start from the Hamiltonian describing the combined system of $\Lambda$-type 3LS and photonic reservoir, cf. Eq.~\eqref{H}.
To facilitate further treatment, we transform this Hamiltonian into the rotating frame defined by its freely evolving part. Together with the rotating-wave approximation the transformation yields 
\begin{equation}
H'(t) = \frac{\hbar \Omega(t)}{2} \left( \sigma_{12} + \sigma_{21} \right)
+  \int \mathrm{d}k \left[ g_k e^{i \omega_k t} r_k^\dagger e^{- i \omega_3  t}  \left(   \sigma_{13} + \sigma_{23} e^{i \omega_2 t}\right) + H.c. \right]. \label{Hprime}
\end{equation}
If the amplitude of the pump laser is constant, i.e. if we are considering a continuous-wave pump field $\Omega$, a transformation into the interaction picture defined by the pump term, that is the first line in Eq.~\eqref{Hprime}, allows further analytical treatment and provides insight into the underlying feedback mechanism. Making use of the specific form of the commutation relation of the atomic flip operators, the transformed Hamiltonian reads
\begin{multline}
H''(t) =  \int \mathrm{d}k \left\{  g_k e^{i \omega_k t}  r_k^\dagger e^{-i \omega_3 t}  \left[ \cos\left(\frac{\Omega}{2} t \right) \left( \sigma_{13} + \sigma_{23} e^{i \omega_2 t}\right) + i \sin\left( \frac{\Omega}{2}t\right) \left( \sigma_{13} e^{i \omega_2 t} + \sigma_{23} \right)  \right] \right. \\
\left. + H.c. \vphantom{\left( \frac{\Omega}{2}t\right)} \right\}.
\label{Htilde}
\end{multline}
We derive equations of motion for the state of the system to be able to study its time evolution. The general state in the single-excitation limit and the respective interaction picture is given as
\begin{equation}
\ket{\psi''(t)} = c_3(t) \ket{3, \text{vac}} + \int\mathrm{d}k c_2^k(t) \ket{2,k} + \int\mathrm{d}k c_1^k(t) \ket{1,k}
\label{state}
\end{equation}
where $\ket{3, \text{vac}}$ describes the case in which the 3LS is in the excited third state and there are no photons in the reservoir while for $\ket{2,k}$ ($\ket{1,k}$) the 3LS is found in state $\ket{2}$ ($\ket{1}$) and there is a photon in reservoir mode $k$. 
Using the Schr\"odinger equation in the interaction picture
\begin{equation}
i \hbar \frac{\mathrm{d}}{\mathrm{d}t} \ket{\psi'' (t)} = H''(t) \ket{\psi'' (t)}, 
\end{equation}
the equations of motion which can be derived for the coefficients of the state given in Eq.~\eqref{state} are
\begin{align}
&\dot{c}_3(t) = - \frac{i}{\hbar}  \int \mathrm{d}k g_k   e^{- i \omega_k t} e^{i \omega_3  t} \notag \\
& \times \left\{ c_2^k(t) \left[\cos\left( \frac{\Omega}{2}t\right) e^{-i\omega_2 t} - i \sin\left(\frac{\Omega}{2}t\right)\right] 
+ c_1^k(t) \left[ \cos\left(\frac{\Omega}{2}t \right) - i \sin\left(\frac{\Omega}{2}t \right) e^{-i \omega_2 t}  \right]\right\}, \label{c3Pump}\\
&\dot{c}^k_2(t) = - \frac{i}{\hbar} g_k e^{i \omega_k t}  e^{-i \omega_3  t} c_3(t) \left[ \cos\left( \frac{\Omega}{2}t\right) e^{i \omega_2 t} + i \sin\left(\frac{\Omega}{2}t \right)\right], \label{c2kPump}\\
&\dot{c}^k_1(t) = - \frac{i}{\hbar} g_k e^{i \omega_k t}  e^{-i \omega_3  t} c_3(t) \left[ \cos\left( \frac{\Omega}{2}t\right) + i \sin\left(\frac{\Omega}{2}t \right) e^{i \omega_2 t} \right]. \label{c1kPump}
\end{align}

We start with an excited 3LS and no photons in the reservoir. That is, we choose the initial conditions $c_3(0) = 1$, $c_2^k(0) = c_1^k(0) = 0$ for all photon modes $k$.
Under this assumption we formally integrate Eqs.~\eqref{c2kPump} and \eqref{c1kPump} and plug the solution into Eq.~\eqref{c3Pump}. After the insertion of the mode-dependent coupling strength $g_k$ encoding the feedback mechanism we finally obtain the delay differential equation for the probability amplitude of the excited state
\begin{widetext}
\begin{multline}
\dot{c}_3(t) = - 2 \Gamma c_3(t) + \Gamma e^{i \omega_3 \tau} \left[\cos\left( \frac{\Omega}{2}\tau\right)\left(1+e^{-i\omega_2 \tau} \right) - i \sin \left( \frac{\Omega}{2}\tau \right) \left( e^{-i \omega_2 t }+ e^{i \omega_2 (t-\tau)}\right) \right]\\ \times c_3(t- \tau)  \Theta(t-\tau). \label{DDEPump2}
\end{multline}
\end{widetext}

Based on Eq.~\eqref{DDEPump2}, we now examine different scenarios which occur for the 3LS depending on the system parameters.

\subsection{Unpumped System}

We first consider the unpumped case, that is $\Omega = 0$, in which the system obeys the equation
\begin{equation}
\dot{c}_3(t) = - 2 \Gamma c_3(t) + \Gamma e^{i \omega_3 \tau} \left(1+e^{-i\omega_2 \tau} \right)  c_3(t- \tau) \Theta(t-\tau) . \label{DDE3LS}
\end{equation}
Without feedback, we would only obtain the first term on the right hand side of Eq.~\eqref{DDE3LS} and the excitation would decay exponentially. With feedback, however, if it holds that $n/\omega_2 = n'/\omega_3$, $n, n' \in \mathbb{N}$, at the delay time
\begin{equation}
\tau = \frac{2 \pi n}{\omega_2} =  \frac{2\pi n'}{\omega_3}
\label{tauUnpumped}
\end{equation}
stabilization is achieved which manifests as continuous-mode excitation trapping \cite{Nemet2019_arxiv}.
This can be understood looking at Eq.~\eqref{DDE3LS}. If for the feedback phase $\phi_2 \equiv \omega_2 \tau$ the condition $\phi_2 = 2\pi n$, $n \in \mathbb{N}$, is fulfilled or equivalently $e^{-i\omega_2 \tau} = 1$, the 3LS acts as an effective 2LS and shows the Pyragas-type feedback-control discussed in Sec. \ref{2LS}. In this case, the rate of the excitation decay is determined by the feedback phase $\phi_3 \equiv \omega_3 \tau$. If $\phi_3 = 2 \pi n'$, $n' \in \mathbb{N}$, i.e. $e^{i \omega_3 \tau} = 1$, the control signal vanishes as soon as $c_3(t) = c_3(t-\tau)$ and the excitation is stabilized. Thus, for a given 3LS with fixed transition frequencies the choice of a suitable delay time $\tau$ leads to the stopping of the excitation decay after a transient period. There are, however, combinations of the transition frequencies where the delay time needed for a stabilization of the excitation lies outside the experimentally accessible range of parameters.

\begin{figure*}
\centering
\includegraphics[width=\linewidth]{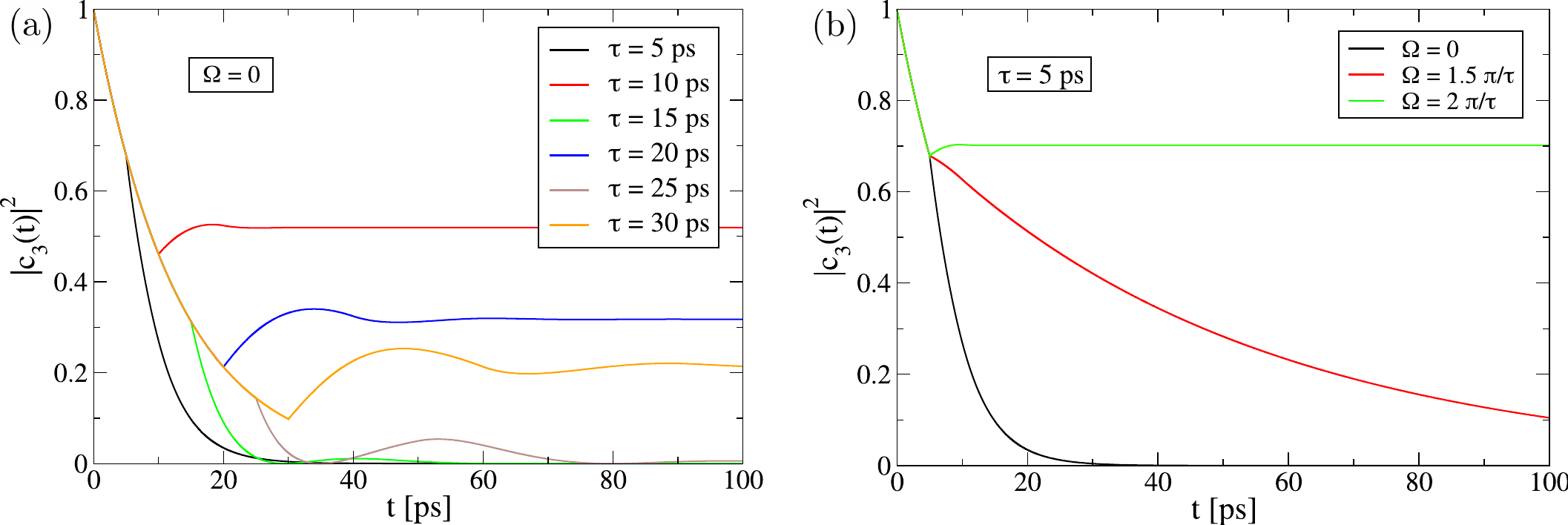}
\caption{Numerical results for the evolution of the excitation probability density $\left| c_3(t)\right|^2$ of the three-level system with $\omega_2/(2\pi) = 0.8$\,$\text{ps}^{-1}$, $\omega_3/(2\pi) = 239.3$\,$\text{ps}^{-1}$, and $\Gamma = 0.01935$\,$\text{ps}^{-1}$. (a)~Unpumped system, i.e. $\Omega=0$, subjected to feedback at different delay times $\tau$. (b)~System subjected to feedback at $\tau = 5$\,ps to which pump fields with different Rabi frequencies $\Omega$ are applied.}
\label{GammaSmall}
\end{figure*}

\begin{figure*}
\centering
\includegraphics[width=\linewidth]{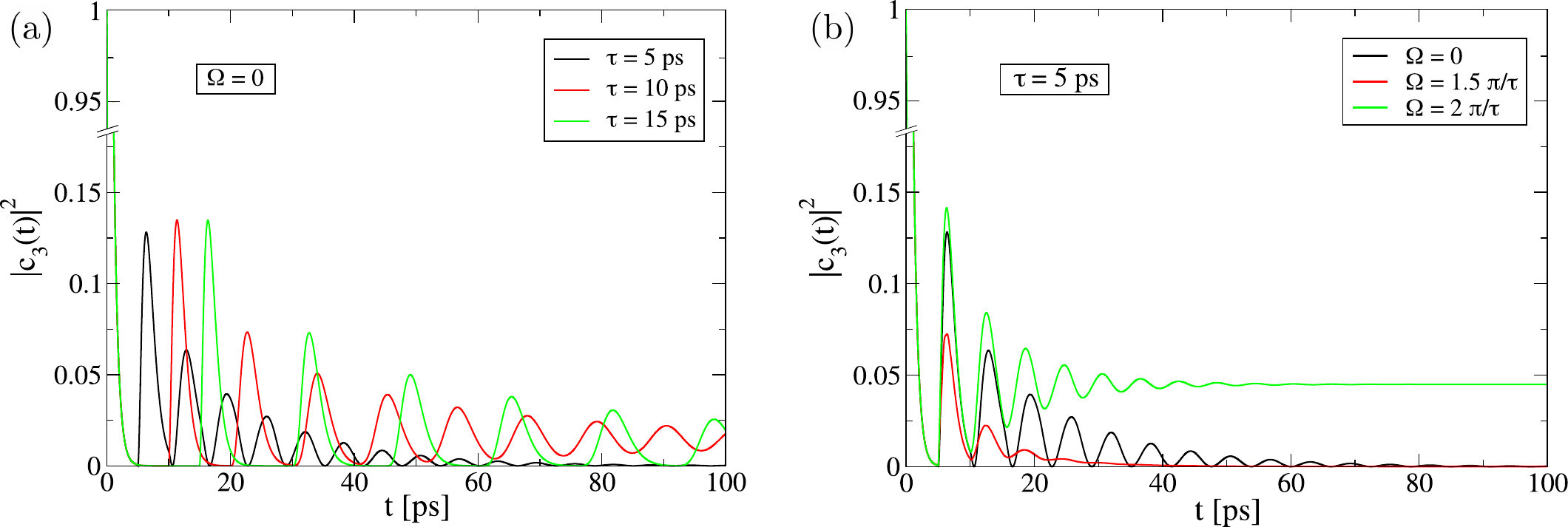}
\caption{Numerical results for the evolution of the excitation probability density $\left| c_3(t)\right|^2$ of the three-level system with $\omega_2/(2\pi) = 0.8$\,$\text{ps}^{-1}$, $\omega_3/(2\pi) = 239.3$\,$\text{ps}^{-1}$, and $\Gamma = 0.37037$\,$\text{ps}^{-1}$. (a)~Unpumped system, i.e. $\Omega=0$, subjected to feedback at different delay times $\tau$. (b)~System subjected to feedback at $\tau = 5$\,ps to which pump fields with different Rabi frequencies $\Omega$ are applied.}
\label{GammaLarge}
\end{figure*}

In Figure~\ref{GammaSmall}~(a), the time evolution of the excitation in a system characterized by $\omega_2/(2\pi) = 0.8$\,$\text{ps}^{-1}$, $\omega_3/(2\pi) = 239.3$\,$\text{ps}^{-1}$ is illustrated. It is subjected to feedback at different delay times $\tau$. We obtain the results from the integration of Eq.~\eqref{DDE3LS} using a Runge-Kutta algorithm.
Because of the 2$\pi$-periodicity of the function $e^{i\phi}$, stabilization is achieved periodically as $\tau$ is varied. If, however, a feedback time not fulfilling the condition given in Eq.~\eqref{tauUnpumped} is chosen, the excitation decays.
The longer the delay time, the more time there is for the excitation to decay before the feedback mechanism acts on the system. As a consequence, the value of the excitation probability density $\left|c_3(t)\right|^2$ at which the system is stabilized if the necessary conditions are fulfilled becomes smaller with increasing $\tau$.
This plays a particularly important role if the system has a a large decay rate $\Gamma$ in comparison to the delay time $\tau$ as is the case in Figure~\ref{GammaLarge}~(a) where the considered 3LS is characterized by the same transition frequencies as the system in Figure~\ref{GammaSmall}~(a) but has a decay rate with  $\Gamma\tau\gg1$. In this system, the excitation has completely decayed before it is fed back into the system by the mirror. For a delay time $\tau = 10$\,ps the excitation is stabilized. There is, however, a transient time of more than ten feedback intervals before a constant value of $\left|c_3(t)\right|^2 = 0.0141$ is reached. For multiples of this delay time, the system can also be stabilized but with an even longer transient time and at a lower value of $\left|c_3(t)\right|^2$.

\subsection{Pumped System}

Next, we turn to the case of a pumped system in which $\Omega \neq 0$. We obtain an additional set of parameters for which the stabilization of the excitation is possible. If the system satisfies the condition
$n/\omega_2 = (n'+1/2)/\omega_3$, $n, n' \in \mathbb{N}$, at the delay time
\begin{equation}
    \tau = \frac{2 \pi n}{\omega_2} = \frac{(2n'+1)\pi}{\omega_3}
\end{equation}
stabilization can be induced using an external pump field.
In the unpumped case, this is the delay time for which the fastest possible decay takes place. If we, however, apply an additional pump field satisfying $\Omega \tau/2 = \pi$, the excitation decay is stopped and the system is stabilized as can be seen from Eq.~\eqref{DDEPump}.

In Figure~\ref{GammaSmall}~(b), the time evolution of the excitation in the same system as in Figure~\ref{GammaSmall}~(a) at a fixed delay time $\tau = 5$\,ps is illustrated. A continuous-wave excitation with different Rabi frequencies $\Omega$ is applied to the system. With increasing values of the Rabi frequency the excitation decay is slowed down and eventually for $\Omega = 2 \pi/\tau$ the decay is stopped. Thus, microwave light fields which are weak compared to the splitting of the energy levels suffice to control the system.
Because of the 2$\pi$-periodicity of the sine and cosine function in Eq.~\eqref{DDEPump}, the excitation again decays if the Rabi frequency is further increased and for all pump fields satisfying $\Omega \tau /2 = (2n''+1)\pi$, $n'' \in \mathbb{N}$, the excitation is stabilized.

In Figure~\ref{GammaLarge}~(b), the system from Figure~\ref{GammaLarge}~(a) is subjected to feedback at $\tau = 5$\,ps. Additionally, pump fields with different Rabi frequencies $\Omega$ are applied. The pump strength $\Omega = 2 \pi / \tau$ leads to a stabilization of the excitation. 
Compared to the stabilization via the tuning of the delay time as presented in Figure~\ref{GammaLarge}~(a) the stopping of the excitation decay via an external pump field shows striking differences:
The transient time is shortened significantly as the constant value of the excitation probability density is already reached after around 50\,ps which corresponds to ten feedback intervals as opposed to the unpumped case in which stabilization is only achieved after more than 100\,ps and ten feedback intervals. Furthermore, stabilization is possible at a shorter delay time $\tau$ so that the amount of trapped excitation $\left|c_3(t)\right|^2 = 0.0449$ for $t \rightarrow \infty$ lies substantially above the one of the unpumped case. It depends on the relation of the transition frequencies $\omega_2$ and $\omega_3$ whether at the shortest delay time, for which it is possible to stop the excitation decay, stabilization is induced with or without an external pump field. If it holds that $\omega_2/\omega_3 = n/n'$, $n, n' \in \mathbb{N}$ and the fraction $n/n'$ is irreducible, we can deduce from the value of $n$ which scenario occurs: If $n$ is even, an external pump field gives rise to excitation trapping at the smallest possible value of $\tau$.
If, on the contrary, $n$ is odd, the system is stabilized at the smallest value of $\tau$ without an additional laser field. In the example presented in Figs.~\ref{GammaSmall} and \ref{GammaLarge} with $\omega_2 /\omega_3 = 8/2393 = n/n'$, $n$ has an even parity. As we have seen above, in this case it is possible to stabilize the system at a higher value of $\left|c_3(t)\right|^2$ using an external pump field than in the unpumped case. In the opposite case of odd parity, however, our approach can also be useful if the minimal delay time $\tau$ at which stabilization is possible without an additional pump field is not accessible due to experimental limitations.

\section{Conclusion and Outlook}
We have discussed the coherent control of quantum few-level systems using time-delayed feedback.
As a possibility to go beyond emerging limitations of conventional schemes of this type of closed-loop control, we have proposed a novel approach to non-invasively control the feedback phase in quantum few-level systems using an external pump laser. As an exemplary system, we introduced the $\Lambda$-type 3LS subjected to feedback for which the application of a resonant pump field between its non-degenerate ground states gives rise to a new control parameter, namely the Rabi frequency $\Omega$ of the pump laser. 

To illustrate the benefits of this approach we presented three examples of conventional coherent feedback control schemes. Thereby, we identified the amplitude and the phase $\phi$ of the feedback signal as being responsible for the influence of the feedback since both affect the way the emitted and the absorbed light interfere. In the considered setups the feedback phase is determined by the delay time and the characteristic system frequency, that is the transition frequency of the respective few-level system or the frequency of the cavity mode depending on the specific setup.

We studied the differences of the control of the 3LS via changes in the delay time and via an external pump field which disentangles the control of the feedback phase from the control of the amplitude.
Compared to the control via the delay time, an external pump field potentially shortens transient times significantly and provides more experimental freedom.
Thus, our approach paves the way for efficient coherent control schemes of quantum few-level systems 
and allows to tailor photon statistics and entanglement properties using time-dependent quantum control. As an outlook, it will be interesting to investigate the possibility of a time-dependent phase and its impact on two- or three photon-dynamics.

\section*{Acknowledgements}
The authors gratefully acknowledge the support of the Deutsche Forschungsgemeinschaft (DFG) through the project B1 of the SFB 910 and from the European Unions Horizon 2020 research and innovation program under the SONAR grant Agreement No. 734690. We thank N. Nemet and J.P. Neudeck for fruitful discussions in the early stages of the project.

\bibliography{MyCollection}

\end{document}